# Polygonal tessellations as predictive models of molecular monolayers

Krisztina Regős,*[a] Rémy Pawlak,[b] Xing Wang,[c] Ernst Meyer,[b] Silvio Decurtins,[c] Gábor Domokos,[a] Kostya S. Novoselov,[d] Shi-Xia Liu,*[c] and Ulrich Aschauer*[c]

a) Department of Morphology and Geometric Modeling, and MTA-BME Morphodynamics Research Group, Budapest University of Technology and Economics, H-1111, Budapest, Hungary.
b) Department of Physics, University of Basel, Klingelbergstrasse 82, 4056 Basel, Switzerland.
c) Department of Chemistry, Biochemistry and Pharmacy, University of Bern, Freiestrasse 3, 3012 Bern, Switzerland.
d) Institute for Functional Intelligent Materials, National University of Singapore, 117544, Singapore.

**Molecular self-assembly plays a very important role in various aspects of technology as well as in biological systems[1–4]. Governed by the covalent, hydrogen or van der Waals interactions - self-assembly of alike molecules results in a large variety of complex patterns even in two dimensions (2D). Prediction of pattern formation for 2D molecular networks is extremely important, though very challenging, and so far, relied on computationally involved approaches such as density functional theory, classical molecular dynamics, Monte Carlo, or machine learning[5–8]. Such methods, however, do not guarantee that all possible patterns will be considered and often rely on intuition. Here we introduce a much simpler, though rigorous, hierarchical geometric model founded on the mean-field theory of 2D polygonal tessellations [9,10] to predict extended network patterns based on molecular-level information. Based on graph theory, this approach yields pattern classification and pattern prediction within well-defined ranges. When applied to existing experimental data, our model provides an entirely new view of self-assembled molecular patterns, leading to interesting predictions on admissible patterns and potential additional phases. While developed for hydrogen-bonded systems, an extension to covalently bonded graphene-derived materials[11,12] or 3D structures such as fullerenes is possible, significantly opening the range of potential future applications.**

Two-dimensional natural patterns at all scales, ranging from molecular assemblies[13,14] to macroscopic entities[15,16], often appear as polygonal tessellations[17]. The corresponding mathematical theory[9,10] could be harnessed to obtain a deeper understanding of the structure-forming process. Our aim here is to make this step for molecular monolayers. Supramolecular tessellation in these materials is often based on non-covalent van der Waals interactions or halogen- and hydrogen bonding between neighbouring molecules.[1–4,18] Illustrative examples of the former interactions are tessellations via exo-wall contacts between shape-persistent polygonal macrocycles, such as pillar[6]arene, which has a hexagonal configuration leading to a trivalent vertex.[19] Similarly, 2D layered network superstructures have been formed using pagoda[4]arene with square and rhombic tiles.[20] Supramolecular tessellation studies based on directional hydrogen bonding are numerous, as the extended structures can be varied over a wide range by careful design of the molecular building blocks.[21–23] Although in recent decades empirical insights into the molecular self-assembly patterns are increasingly obtained from experimental scanning tunnelling microscopy (STM) data at molecular resolution that often allow for an intuitive assessment of the resulting molecular tiles, theoretical methods leading to a deeper understanding and predictability of pattern formation are still urgently needed.

Theoretical investigations have shown the architecture and symmetry of the molecular precursor to play an important role in combination with the nature of the intermolecular interactions.[7] In a recent study, Baran et al. investigated how the two important factors of complementary molecular size and shape in combination with the presence of directed interactions control self-assembly at the surface.[24,25] Using coarse-grained modelling, their study aims to both reproduce experimental data and predict new supramolecular structures, thereby gaining an understanding of the factors playing a key role in the self-assembly process. In addition, molecular dynamics and related computational approaches have been applied to various molecular architectures[5,6]. Despite continuous advances in efficient algorithms and computational speed, reorientation kinetics are often too slow to capture all possible configurations and reliably identify the most stable structure. Enhanced sampling and Monte Carlo[6] or machine learning[7,8] approaches have alleviated this issue but still require significant computational resources.

In this work, we present an alternative theoretical approach based on the mean-field theory of space-filling polygonal mosaics[9,26,27] which describes these structures by average values $\bar{n}$, $\bar{v}$ of the nodal and cell degrees. The former counts the number of polygons meeting at one vertex, and the latter counts the number of vertices of a polygon (see Fig. 1). The averages $\bar{n}$, $\bar{v}$ span the *symbolic plane* and each space-filling mosaic is associated with one point on this plane. This representation proved useful in determining the provenance of natural fragmentation patterns[15,16]. As compared to fracture patterns, molecular assemblies show special features: if they are space-filling, they are also *regular* (i.e., the vertex of one polygon is not admitted lying in the interior of an edge of another polygon) in which case the averages are not independent:[26,27]

$$\bar{v} = \frac{2\bar{n}}{\bar{n}-2}, \quad (1)$$

therefore, if one of the averages can be predicted, the other may be computed using (1).

Single-layer materials may be represented by a hierarchical geometric model, consisting of (parts of) polygonal mosaics on multiple scales, which we dub *Levels* and denote, in hierarchical order, by $L_i$, (i=1,2,3,4), associated with the averages $\bar{n}_i$, $\bar{v}_i$. We note that the concept of levels was first introduced to describe the fractal geometry of 2D foams[28]. Our approach is conceptually similar in the sense that scales of subsequent levels are well separated, however, in our case the subsequent levels do not carry self-similar patterns. Instead of being interpreted as convex mosaics, levels may also be viewed as *graph representations* of the monolayer. In this context Levels $L_2$ and $L_4$ emerge as the two *essential models*, being, respectively, the *fully expanded* and the *fully contracted graphs* associated with the monolayer. We next illustrate these levels on the example of the α-phase of the 2,7-pyrenedione (PO) molecule (see SI Section S1.1) as shown in Fig. 2.

Level $L_1$ is not space-filling and describes the geometry of the interacting sites of a single molecule as a convex polygon with $\bar{v}_1$ vertices at the molecule's *perimeter atoms* (i.e., the atoms able to form intermolecular bonds). For the PO example we have $\bar{v}_1 = 10$ (8 hydrogen and 2 oxygen atoms). In addition, we also characterize the chemical bonding of the molecule via the parameters $b_1$ and $\hat{b}_1$. For the hydrogen-bonded systems we focus on, $b_1$ is the *maximum acceptor capacity* of any atom among the $\bar{v}_1$ perimeter atoms. For PO $b_1 = 2$, i.e., the maximum number of accepted hydrogen bonds for the oxygen atom. Based on the molecule's total number of accepted ($b_a$) and donated ($b_d$) hydrogen bonds, we define $\hat{b}_1 = \min(b_a, b_d)$. For PO $b_a = 4$ (two oxygen atoms with two accepted bonds each) and $b_d = 8$ (8 hydrogen atoms with one donated bond each), so $\hat{b}_1 = \min(4,8) = 4$ (see Figure S5).

Levels $L_2$, $L_3$, and $L_4$ describe supramolecular space-filling tessellations, so formula (1) can be used and in each case the single average $\bar{n}_i$ characterizes the pattern. As noted, Level $L_2$ is an essential model characterized by the fully expanded graph: here *all* perimeter atoms appear as nodes and besides the edges of the $L_1$–polygons (molecular perimeter) *all* intermolecular bonds appear as edges (see Figure 2). On level $L_2$ we also define $b_2$ as the average number of bonds between neighbouring molecules. For the PO α-phase the nodal and cell degree averages are $(\bar{n}_2, \bar{v}_2)$ = (2.633, 8.316), and the average bonding number $b_2 = 1.357$. We derive the graph corresponding to Level $L_3$ from level $L_2$ via *chemically targeted* face contractions: all faces of the $L_2$ graph (i.e., cells of the $L_2$ mosaic) corresponding to molecules are contracted to single nodes (see Figure 2), such that, on level $L_3$ the nodes are the molecules, and each intermolecular bond is counted as an edge. For the PO α-phase we have nodal and cell degree averages $(\bar{n}_3, \bar{v}_3)$ = (6.333, 2.923), Level $L_4$, finally, is again an essential model, characterized by the fully contracted graph. The latter is, as before, obtained from the preceding level $L_3$ via *chemically targeted* face contractions: all faces of the $L_3$ graph (cells of the $L_3$ mosaic), the perimeter of which is formed by two intermolecular bonds, are contracted to single edges (see Figure 2), so, on level $L_4$ the nodes are the molecules, and a single edge is added for any two bonded molecules, independent of the number of bonds they share. For the PO α-phase $(\bar{n}_4, \bar{v}_4)$ = (4.667, 3.500).

These definitions can be used to organize supramolecular patterns using the symbolic plane. Fig. 3 shows the Level $L_2$ patterns associated with some 2D materials (to be discussed in more detail below and in the Supporting Information (SI) section S1) as well as some simple periodic patterns in the symbolic plane along the curve defined by equation (1).

The model can be made predictive by expressing the supramolecular pattern averages $\bar{n}_2, \bar{n}_3, \bar{n}_4$ by the molecular pattern average $\bar{v}_1$ and the chemical bonding information carried by the constants $b_1, \hat{b}_1$ and $b_2$. While the derivation can be found in the SI section S2, we here present the formulae along with some physical interpretation. Formulae (2), (3), and (4) determine *higher-level* geometric information based on *lower-level* geometric and chemical information:

$$\frac{2}{\bar{v}_1} + 2 \leq \bar{n}_2 \leq \frac{2\hat{b}_1}{\bar{v}_1} + 2 \leq \frac{2b_1}{b_1+1} + 2 \quad (2)$$

$$\bar{n}_3 = \bar{v}_1(\bar{n}_2 - 2) \quad (3)$$

$$\bar{n}_4 = \frac{\bar{n}_3}{b_2} \qquad (4)$$

In addition, the *mixed* formula (5), expresses mean-field averages using both lower-level and higher-level information:

$$\frac{\bar{n}_4}{\bar{v}_1} + 2 \leq \bar{n}_2 \qquad (5)$$

We note that if the level 4 pattern is convex then in formula (5) $\bar{n}_4 \geq 3$. The geometric model formed by formulae (2)-(5) was found to be correct for the experimental data of 2D ice as well as PO, TAPE, and TT molecules (see data points in Fig. 3 and SI section S1 for numeric values as well as a description of the molecules). We note that formula (2), and particularly the right-most upper bound predicts that $\bar{n}_2 \leq 4$ and consequently $\bar{v}_2 \geq 4$ for all level $L_2$ supramolecular patterns. This implies for all conceivable supramolecular patterns $\bar{n}_2 \leq \bar{v}_2$, thus lying towards the left on the curve given by formula (1), greatly restricting the variety of possible patterns. In addition, we found that the bounds given by equations (2) and (5) are sharp, i.e., there exist known 2D molecular monolayers where these bounds are exactly realized. This can be seen for the case of 2D ice in Fig. 3, where the hexagonal and square arrangements lie on the upper and lower bounds respectively of the admissible range (shown by the red arrow) determined by a combination of formula (5) for the lower and formula (2) for the upper bound:

$$\frac{3}{\bar{v}_1} + 2 \leq \bar{n}_2 \leq \frac{2\hat{b}_1}{\bar{v}_1} + 2 \qquad (6)$$

In formula (5) we used the information that the level $L_4$ pattern is convex. This implies that these bounds cannot be improved unless one imposes restrictions on the set of considered materials.

One notable feature of the admissible ranges for water and PO shown in Fig. 3 is that they do not overlap on either axis, implying that the two molecules will never produce identical supramolecular patterns.

Apart from the PO α-phase, mentioned in deriving the levels, the same molecule may also form a β-phase, which is characterized by a slightly larger $\bar{n}_2$. Neither of the two phases, however, lies on the boundary of the admissible range (shown by the purple arrow), implying that further phases could exist. While no structure outside the range delimited by the α- and β-phase was observed experimentally, a γ-phase falling within this range was observed, which has $(\bar{n}_4, \bar{v}_4)$ = (3.600, 4.500). We observe a potentially interesting symmetry at level $L_4$: the α- and β-phases correspond to the symmetrical, dual points $(\bar{n}_4, \bar{v}_4)$ at (4.667, 3.500) and (3.500, 4.667) respectively (see SI section S1). The underlying physical meaning of this symmetry uncovered by our mean-field geometric model and its validity for other molecular architectures will be an interesting topic for future research.

A further interesting aspect is that the model may be tuned by adding additional data to the chemical bonding information. Temperature influences the bonding constant $\hat{b}_1$ and, via equation (2), also the combinatorial averages. For PO, $\hat{b}_1$ has been computed via Boltzmann populations based on DFT total energies of single and bifurcate hydrogen bonds (see details in SI section S3). As the preference for bifurcate hydrogen bonds is reduced with increasing temperature, $\hat{b}_1$ decreases and the lower bound for $\bar{v}_2$ increases as a function of temperature (see SI Fig. S4). This implies that at finite temperature the admissible range for patterns may shrink.

In summary, we have applied the theory of space-filling polygonal tessellations to classify 2D molecular monolayers and obtained a predictive understanding of the conceivable geometries. While developed here for hydrogen-bonded molecular networks, there is no fundamental reason restricting our method to this class of materials. Indeed, only the higher upper bound in equation (2) is directly related to hydrogen bonding, while the rest of the model could be readily applied to other 2D assemblies, for example covalently bonded atomic systems like the rich class of graphene derived compounds such as porous and doped graphene[29] or related compounds such as boron nitrate[11] or borophene[12]. As the concept of the $(\bar{n}, \bar{v})$ symbolic plane does not depend on the dimension of the tessellation, the same concepts could, in principle, be applied to the classification of 3D supramolecular patterns. However, in 3D no formula analogous to (1) exists, so we expect that patterns would occupy not just a curve, but a domain of the symbolic plane. Our theory has shown its promise for the classification and separation of 2D molecular networks. Given that our approach yields a set of numbers classifying any given molecule and assembly, it could be of high relevance in machine-learning studies of the temperature-dependent relation between molecular structure and the resulting assembly, serving as a pre-processing step to turn image data into suitable numerical datasets.

**Acknowledgements**

Aisha Ahsan and Thomas A. Jung are gratefully acknowledged for access to unpublished data. K.R. and G.D. were supported by the Hungarian Research Fund (NKFIH) grant 134199 and NKFIH Fund TKP2021 BME-NVA. K.R. received further support from the Program ÚNKP-22-3 by the Hungarian Ministry of Innovation and Technology (ITM) and NKFIH and gratefully appreciates the gift representing the Albrecht Science Fellowship. S.-X.L. was supported by the Swiss National Science Foundation (SNF) grant 200021_ 204053. X.W. and U.A. acknowledge funding by the SNF Professorship PP00P2_187185/2 and Project 200021_178791. DFT calculations were performed on UBELIX (http://www.id.unibe.ch/hpc), the HPC cluster at the University of Bern.

**Author Contributions**

K.R., S.D., G.D., K.S.N., S.-X.L., and U.A. conceptualized the research. K.R. and G.D. developed the geometric model. X.W. and U.A. performed DFT calculations. R.P., and E.M. provided unpublished experimental data. All authors wrote and edited the manuscript.

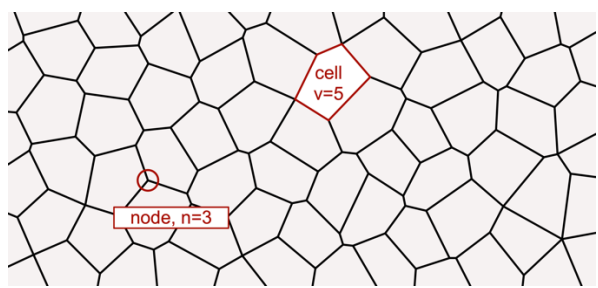

**Fig. 1** Regular polygonal mosaic in the plane. The cell degree $v$ counts the vertices of a polygon, the nodal degree $n$ counts the polygons overlapping at a vertex.

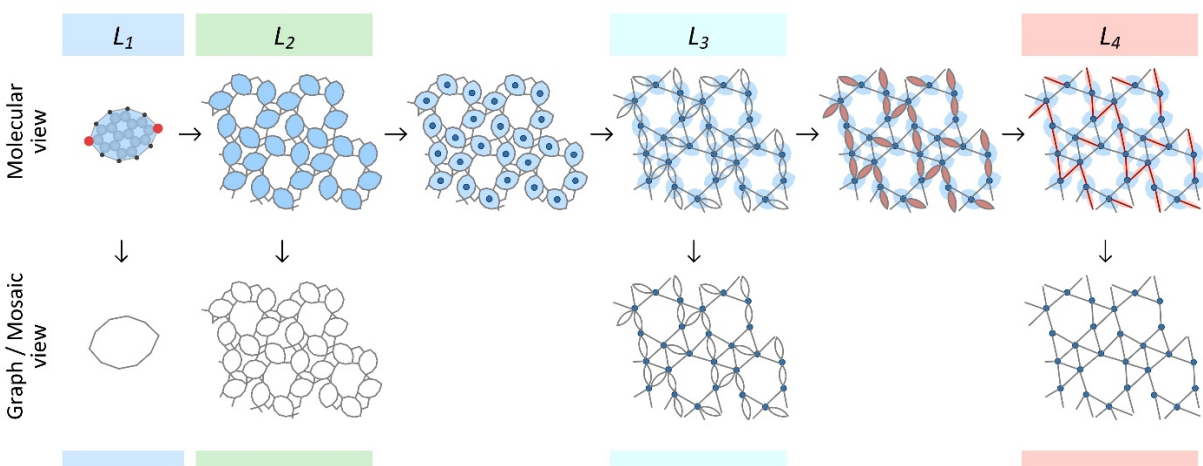

**Fig. 2** Illustration of the Levels as an interpretation of molecular patterns, using the α-phase of the 2,7-pyrenedione (PO) molecule. Upper row: physical images of molecules and patterns. Lower row: abstract graph/mosaic representation of patterns. Columns: Level $L_1$: Structure of the molecule and blue polygon derived from the perimeter atoms (upper row), perimeter as convex polygon (lower row). Level $L_2$: Supramolecular pattern. Faces corresponding to molecules are filled with blue colour (upper row) and shown in graph representation (lower row). Level $L_2 \rightarrow L_3$ transition shown in the upper row: midpoint of blue molecular faces marked as new nodes. Level $L_3$: Pattern obtained by contracting blue molecular faces to single nodes. Lower row: graph of Level $L_3$. Level $L_3 \rightarrow L_4$ transition shown in the upper row: faces between multiple adjacent bonds coloured red. Level $L_4$: Pattern obtained by contracting red faces to edges. Lower row: graph of Level $L_4$.

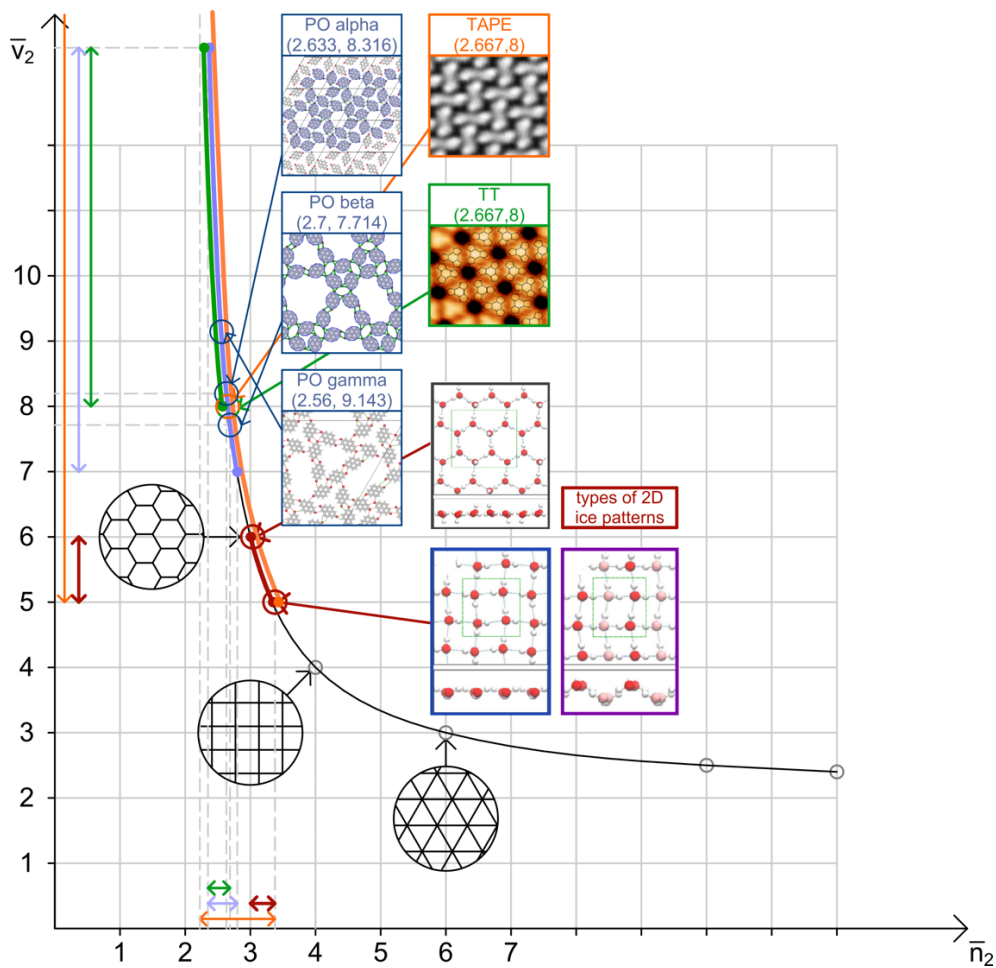

**Fig. 3** Level $L_2$ supramolecular patterns organized along the line of equation (1) in the symbolic plane. 2D ice structures adapted from[30].

# Polygonal tessellations as predictive models of molecular monolayers

Krisztina Regős,*[a] Rémy Pawlak,[b] Xing Wang,[c] Ernst Meyer,[b] Silvio Decurtins,[c] Gábor Domokos,[a] Kostya S. Novoselov,[d] Shi-Xia Liu,*[c] and Ulrich Aschauer*[c]

a. Department of Morphology and Geometric Modeling and MTA-BME Morphodynamics Research Group, Budapest University of Technology and Economics, H-1111, Budapest, Hungary.
b. Department of Physics, University of Basel, Klingelbergstrasse 82, 4056 Basel, Switzerland.
c. Department of Chemistry, Biochemistry and Pharmacy, University of Bern, Freiestrasse 3, 3012 Bern, Switzerland.
d. Institute for Functional Intelligent Materials, National University of Singapore, 117544, Singapore.

## Section S1: Details on material examples

We have collected data on several prototypical molecular examples that exhibit a planar polycyclic structure with embedded heteroatoms that enables the formation of multiple hydrogen-bonds within extended surface patterns.

### Section S1.1: PO

2,7-dihydroxypyrene undergoes thermally activated dehydrogenation to 2,7-pyrenedione (PO) under the experimental protocol of adsorption on an Ag(111) substrate.[1] This planar molecule with $D_{2h}$ symmetry is pre-functionalized to allow multiple C-H⋯O hydrogen bonds in an extended 2D arrangement (Figure S1). Interestingly, different nanoporous network structures, termed α-, β- and γ-phase, coexist on an Ag(111) substrate without mixing.[2] The α phase consists of closed-packed hexagonal arrays with a lattice parameter of 24 Å and identical hexagonal cavities. The β-phase has a hexagonal structure with a lattice parameter of 87 Å and has three-, six-, eight- and nine-membered cavities. The γ-phase has a hexagonal structure with a lattice parameter of 67 Å but only three-, six- and nine-membered cavities. These extended nanoporous structures represent periodic patterns with different confinements of surface electrons, i.e., a 2D quantum dot superlattice.[2]

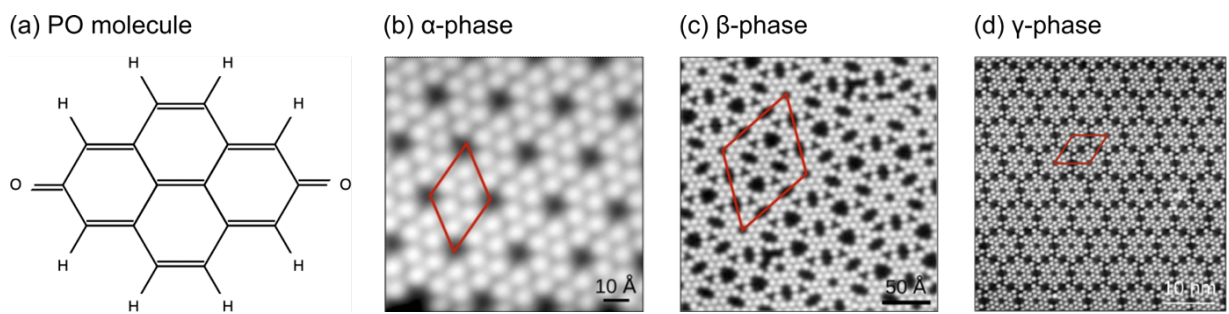

**Figure S1.** (a) The PO molecule with its different C-H⋯O hydrogen bonding capacities. (b-d) STM images of the α-, β-, and γ-phases, respectively, of PO molecules on Ag(111), T = 5 K. Panels (b) and (c) from Ref. [2] with permissions of the authors.

## Section S1.2: TAPE

The planar 1,6,7,12-tetraazaperylene (TAPE) exhibits a $D_{2h}$ symmetry and is characterized by the ability to form multiple intermolecular C-H⋯N hydrogen bonds within a 2D arrangement (Figure S2).[3–5]

(a) TAPE molecules

(b) resulting pattern

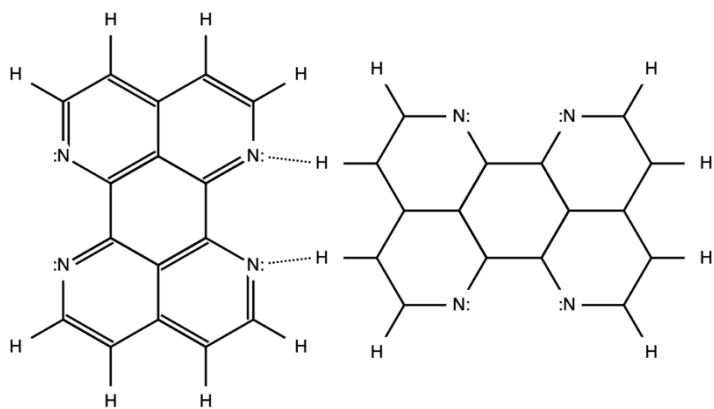
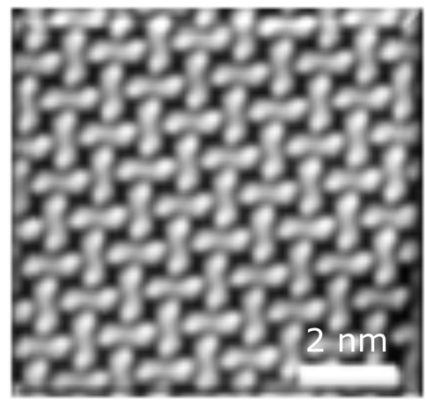

**Figure S2.** (a) Two adjacent TAPE molecules forming intermolecular C-H⋯N hydrogen bonds. (b) STM image of an assembly of TAPE molecules on Ag(111), T = 6 K.

## Section S1.3: TT

The planar triimidazo[1,3,5]triazine (TT) exhibits a $D_{3h}$ symmetry and is pre-functionalized to allow multiple C-H⋯N hydrogen bonds in an extended 2D arrangement (Figure S3).[6]

(a) TT molecules

(b) resulting pattern

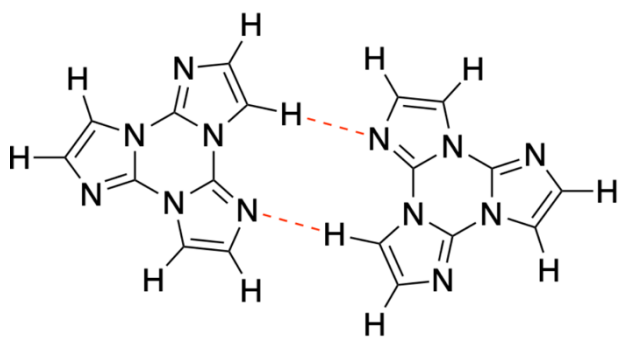
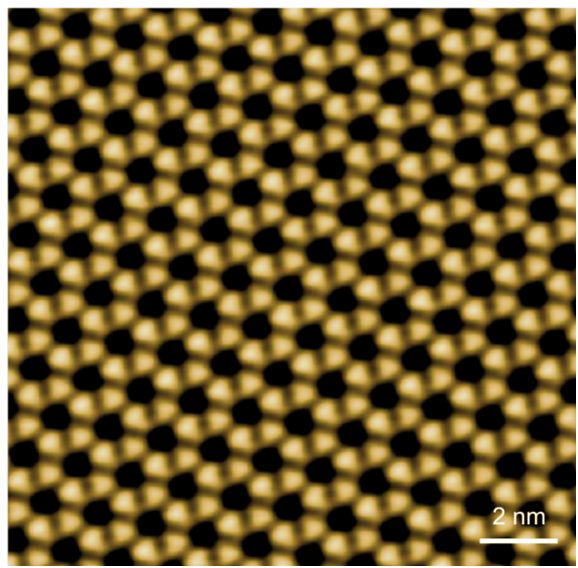

**Figure S3.** (a) Two adjacent TT molecules forming intermolecular C-H⋯N hydrogen bonds and (b) a STM image of a TT assembly on Ag(111).

**Section S1.4: 2D-ice:**

We examined three variants for single-layer water ice assemblies, see panels (a), (c), (d) in Figure 3 of the main text, which were taken from[7], Fig. 1). The mean field averages for the rectangular grids in panels (c) and (d) are straightforward. In the case of the hexagonal pattern in panel (a) the computation of $\bar{n}_2$ is nontrivial, as the level 2 assembly is essentially non-planar. We can proceed in two ways (and we arrive at the same result):

1. We can count the nodal averages directly on the Level 2 assembly, ignoring the fact that some bonds are not in the plane. Here we get $\bar{n}_2 = \frac{8*3+2*4+2*2}{12} = 3$.
2. Alternatively, we can determine $\bar{n}_2$ by using higher and lower-level information, thus combining formulae (3) and (4). Here we start with $\bar{n}_4 = 3, b_2 = 1$, and we get from (4) $\bar{n}_3 = 3$. Substituting $\bar{v}_1 = 3$ into (3) we arrive at $\bar{n}_2 = 3$, as in the previous computation.

Based on these considerations we identified the hexagonal 2D ice pattern with the point $(\bar{n}_2, \bar{v}_2) = (3,6)$ on the symbolic plane.

**Section S1.5: Summary**

Table S1 summarizes the bonding parameters and geometric averages determined for all considered molecules and patterns.

**Table S1.** Parameters and geometric averages of all considered molecules and patters.

|  | PO alpha | PO beta | PO gamma | TAPE | TT | 2D-ice |
|---|---|---|---|---|---|---|
| $\bar{n}_0, \bar{v}_0$ | 3,6 | 3,6 | 3,6 | 2.8,6 | 3,5.25 | - |
| $\bar{n}_1, \bar{v}_1$ | 2,10 | 2,10 | 2,10 | 2,12 | 2,9 | 2,3 |
| $b_a$ | 4 | 4 | 4 | 4 | 3 | 2 |
| $b_d$ | 8 | 8 | 8 | 8 | 6 | 2 |
| $\hat{b}_1$ | 4 | 4 | 4 | 4 | 3 | 2 |
| $b_1$ | 2 | 2 | 2 | 1 | 1 | 2 |
| $\bar{n}_2, \bar{v}_2$ | 2.633,8.316 | 2.7,7.714 | 2.560, 9.143 | 2.667,8 | 2.667,8 | 3.33,5 |
| $\bar{n}_3, \bar{v}_3$ | 6.333,2.923 | 7,2.8 | 3.111, 5.600 | 8,2.667 | 6,3 | 4,4 |
| $\bar{n}_4, \bar{v}_4$ | 4.667,3.5 | 3.5,4.667 | 3.600, 4.500 | 4,4 | 3,6 | 4,4 |
| $b_2$ | 1.357 | 2 | 1.555 | 2 | 2 | 1 |

## Section S2: Derivation of the formulae

### Section S2.1: Derivation of formula (2)

- The nodes of Level 2 are the perimeter atoms. Each of these atoms is connected to two neighbours in the same molecule, this adds 2 to their minimal degree. We also consider that on Level 4 the molecule itself is a node of a space-filling polygonal pattern, and in any such pattern the degree of any node is at least 2. In other words, there are at least two other molecules with which bonds exist, adding $\frac{2}{\bar{v}_1}$ to the minimum of the average nodal degree. This yields $\frac{2}{\bar{v}_1} + 2 \leq \bar{n}_2$.

- Since there is only one molecule in the pattern, all accepted H-bonds must be matched by donated H-bonds. If in one molecule we add up all potential accepting and donating bonds as $b_a, b_d$, the double of the minimum of these two quantities will give an upper bound on the total number of external bonds. To obtain the average we divide by the number of perimeter atoms and add 2 for the connections along the perimeter and we arrive at $\bar{n}_2 \leq \frac{2\hat{b}_1}{\bar{v}_1} + 2$.

- We assume that all donating bonding capacities are 1 (all donating bonds made by Hydrogen atoms). Let us replace the capacity of each atom with accepting bonding capacity (those which are not hydrogen atoms) with the maximum $b_1$. Clearly, if we have $x$ atoms with accepting bonding capacity $b_1$, and all these bonds are formed, there must be $xb_1$ Hydrogen atoms participating in those bonds. Also considering that each atom has 2 connections along the perimeter, the average capacity for all external bonding can be written as

$$\bar{n}_2 \leq 2 + \frac{xb_1 + xb_1}{xb_1 + x} = \frac{2b_1}{b_1 + 1} + 2$$

- Our aim is to prove the last part of formula (2): $\frac{2\hat{b}_1}{\bar{v}_1} + 2 \leq \frac{2b_1}{b_1+1} + 2$. Let us replace all atoms with accepting bonding capacity with the maximum $b_1$ again. If we have x atoms with accepting bonding capacity, and y H atoms, we have $b_a = xb_1$ and $b_d = y$. Then $\hat{b}_1 = \min(xb_1; y)$, and $\bar{v}_1 = x + y$.

  o if $xb_1 \leq y$
  $$\frac{2\hat{b}_1}{\bar{v}_1} + 2 = \frac{2xb_1}{x + y} + 2 \leq \frac{2xb_1}{x + xb_1} + 2 = \frac{2b_1}{1 + b_1} + 2$$

  o if $xb_1 \geq y$
  $$\frac{2\hat{b}_1}{\bar{v}_1} + 2 = \frac{2y}{x + y} + 2 = 2 - \frac{2x}{x + y} + 2 \leq 2 - \frac{2x}{x + xb_1} + 2$$
  $$= \frac{2x + 2xb_1 - 2x}{x + xb_1} + 2 = \frac{2b_1}{1 + b_1} + 2.$$

### Section S2.2: Derivation of formula (3)

By definition, $\bar{n}_3$ counts the average number of intermolecular bonds per molecule. On average, each such atom has $\bar{n}_2$ bonds, out of which 2 do not count because they are not intermolecular bonds (rather, connections along the perimeter of the molecule). So, the average number of intermolecular bonds per atom is $(\bar{n}_2 - 2)$ and we must multiply this by the number $\bar{v}_1$ of perimeter atoms to obtain the correct result for the molecule.

### Section S2.3: Derivation of formula (4)

Level 4 and Level 3 differ because in the former case all bonds between any pair of molecules are counted as one. So, if we divide the nodal degree $\bar{n}_3$ associated with Level 3 by the average number of bonds between pairs of molecules, we get $\bar{n}_4$. However, the average number of bonds between pairs of molecules is given as $b_2$ (as chemical information).

### Section S2.4: Derivation of formula (5)

Each molecule is bonded with at least $\bar{n}_4$ other molecules, so $\frac{\bar{n}_4}{\bar{v}_1}$ provides the minimal average number of external bonds per atom. Since we also have two connections to the perimeter atoms, we can obtain the lower bound for the nodal degree on Level 2 as $\frac{\bar{n}_4}{\bar{v}_1} + 2 \leq \bar{n}_2$.

## Section S3: Thermal effects

Total energies for an isolated PO molecule as well as PO dimers with mono- (one H-bond per O) and bidentate (two H-bonds per O) H-bond configurations were calculated using density functional theory (DFT) as implemented in the Quickstep code[8] of the CP2K package[9] based on a mixed[10] DZVP-MOLOPT-SR[11] Gaussian and plane-wave basis in conjunction with Goedecker, Teter, and Hutter pseudopotentials[12]. The plane-wave cutoff was set to 500 Ry. Exchange and correlation were described by the revPBE[13,14] functional and van der Waals (vdW) interactions were also accounted for[15,16]. Structures were relaxed until forces converged below 0.02 eV/Å.

Based on these energies, the bond energies per H-bonded O were calculated to be 0.18 eV in the monodentate and 0.21 eV in the bidentate configuration (see Fig. S4(a) and (b)). The ratio of O with one vs. O with two bonds was expressed as the temperature-dependent Boltzmann ratio

$$r_{1:2} = exp\left(\frac{E_2 - E_1}{kT}\right)$$

results being shown in Table S2. This implies that, for example at 300 K, 0.259/(1+0.259)=21% of O atoms have 1 H-bond, and 79% have 2 H-bonds. This information can be used to adjust the bonding constant $\hat{b}_1$ and thus affect the allowed range of the combinatorial averages as shown in Fig. S4(c).

**Table S2**. Temperature-dependent Boltzmann ratios

| Temperature | $r_{1:2}$ |
|---|---|
| 10 | 0.000 |
| 100 | 0.017 |
| 200 | 0.132 |
| 300 | 0.259 |
| 400 | 0.364 |
| 500 | 0.445 |

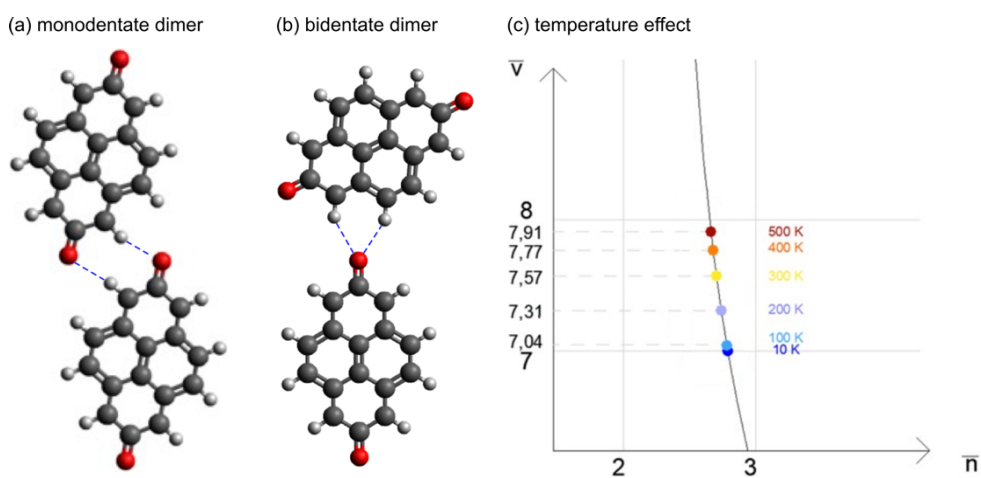

**Figure S4**: (a-b) configurations of PO dimers, (c) thermal effects on the lower $\bar{v}_2$ boundary for the 2,7-pyrenedione (PO) molecule.

**Section S4: Illustration of chemical information in the main formulae**

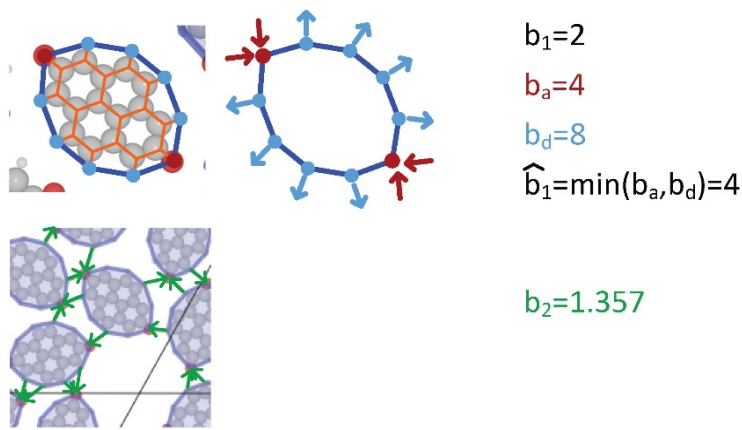

**Figure S5.** Chemical information used in formulae (2)-(4) of the paper, explained on the example of the α-phase of the 2,7-pyrenedione (PO) molecule. Upper image: maximum acceptor capacity $b_1$ of any atom among the perimeter atoms as well as the molecule's total number of accepted ($b_a$) and donated ($b_d$) hydrogen bonds and $\hat{b}_1 = \min(b_a, b_d)$. Lower figure: $b_2$ as the average number of bonds between neighbouring molecules.